 \documentclass[preprint,12pt]{elsarticle}
\usepackage{multirow}
\usepackage{multicol}
\usepackage{amsmath}
\usepackage{amsfonts}
\usepackage{amssymb}
\usepackage{graphicx}
\usepackage{epstopdf}
\usepackage{epsfig}
\usepackage{subfigure}
\usepackage{doi}
\usepackage{hyperref}
\newcommand{\vect}[1]{\boldsymbol{#1}}
\renewcommand{\vec}[1]{\boldsymbol{#1}}

\biboptions{numbers,sort&compress}
\journal{Physics Letters A}

\begin{document}
\begin{frontmatter}

\title{Effect of winding edge currents}

\author[cnrs]{S. Ouvry}
\address[cnrs]{Laboratoire de Physique Th\'eorique et Mod\'eles Statistiques, Universit\'e Paris Sud - CNRS, 91405 Orsay, France}
\author[ltp]{L.  Pastur}
\author[ltp]{A.  Yanovsky}
\address[ltp]{B. Verkin Institute for Low Temperature Physics and Engineering of the National Academy of Sciences of Ukraine, 47 Lenin Avenue, Kharkov 61103, Ukraine}

\begin{abstract}
We discuss  persistent currents for particles with internal
degrees of freedom. The currents arise because of  winding properties essential for the chaotic
motion of the particles in a confined geometry. The currents do not change
the particle concentrations or thermodynamics, similar to the skipping orbits in a magnetic
field.
\end{abstract}


\end{frontmatter}

\section{Introduction}
Windings of random walk trajectories are of considerable
interest in various fields of condensed matter physics, from polymers and
DNA to superconductors and Bose-Einstein condensates (see e.g.
\cite{Ce:95,De-Ou:11,Gr-Fr:03} and references therein). In this paper, we
consider the influence of winding on the internal degrees of freedom of
particles dynamics.

Note that winding dependent degrees of freedom are not
exotic. The most obvious of these is, of course, the spin, which is of
particular interest in view of the recent spintronic surge of activity.
Because of the spin-orbit interaction, the spin
is directly related to the winding of particle trajectories. Accordingly,
our analysis is intimately related to the so-called spin Hall effect in
electron transport \cite{DyakonovSpinHall, HirschSpinHall,
ZhangSpinHall, RashbaExtrSpinHall, DasSarmaExtrSpinHall, SternExtrSpinHall,
JungwirthSpinHallDevices} as well as to other
types of spin transport, like,  for example,  photon (light) propagation in disordered media
(since  photons also have non-zero spin).

Other degrees of freedom, both classical and
quantum, can also be winding  dependent, as e.g.  geometric phases (the Berry
phase). Thus, the results of the present work are applicable not only to  spin
transport but also to any situation that can be reduced to  random trajectories
in a confined geometry with the dependence of  certain characteristics on
winding properties.

The paper is organized as follows. Section 2 presents the concept of
winding and the corresponding terminology. Section 3 deals with the spin
degree of freedom as  directly dependent on a winding angle. Section
4 contains numerical simulations in the case of diffusion in a
rectangular 2d domain with periodic boundary conditions in the longitudinal
direction and hard walls in the transverse direction.  Section 5 proposes a
simple theoretical model of classical random walks with
so-called ``soft walls'', where the geometric constraints are described by an
external confining potential in the transverse direction. Section 6 contains
a brief discussion of the results.

\section{Winding of curved trajectories}

Let us introduce the winding angle $\vec{\theta}$ as the vector sum of all the
turnings of a particle trajectory $\vec{x}(t)$.
The angle is called the \textit{total curvature}
in the literature (see e.g. \cite{Milnor1950,Sullivan2007}). It is given by
the integral of the differential curvature over a natural parameter (the local time
of the trajectory). The differential curvature $\vec{\kappa}$ of the trajectory is
defined as
\begin{equation*}
\vec{\kappa} = |\vec{\dot{x}}|^{-3}\vec{\dot{x}}\times\vec{\ddot{x}},
\end{equation*}
where the  symbol ``$ \,\dot{\;}\, $'' \ denotes  differentiation with respect to $t$ and the  symbol
$\times$ denotes the vector product. Correspondingly, $\vec{\theta}$ has the form
\cite{KuhnelDiffGeometry}
\begin{equation}  \label{eq:thetadef}
\vec{\theta} = \int^t_0 \vec{\kappa}(t)|\vec{\dot{x}}(t)|dt = \int_0^t \frac{%
\vec{v} \times \vec{\dot{v}}}{v^2} dt, \ \vec{v} \equiv \vec{\dot{x}}.
\end{equation}
Note that the total curvature of a closed planar trajectory, equal to the
"winding number" or "turning number" multiplied by $2 \pi$, is a homotopy
invariant.

It is worth mentioning that the term ``\textit{winding angle}'' is also
used in polymer physics (see e.g. \cite{Gr-Fr:03}) where, however,  it rather refers to
\begin{equation*}
\vec{\vartheta} = \int^t_0 \frac{\vec{x} \times \vec{\dot{x}}}{x^2} dt.
\end{equation*}
Thus, $\vec{\vartheta}$ is the winding angle of the radius vector of
the trajectory  in  configuration space, whereas $\vec{\theta}$ yields an
analogous quantity but in  velocity space.

The distribution of $\vec{\vartheta}$'s for  Brownian trajectories  has been
studied for a long time (see e.g. \cite{De-Ou:11,Gr-Fr:03, Spi}), while that
of $\vec{\theta}$'s has not been analyzed so far. On the other
hand, it seems that $\vec{\theta}$ is a more appropriate quantity for certain problems,
for example those including spin (spin Hall effect, etc), as we will argue below.

We consider for simplicity the 2d  case. Generalizations to 3d and tensors
are fairly simple, but the resulting formulas are more involved and less intuitive.
In the planar case the winding angle vector $\vec{\theta}$ is normal to the
plane, hence it suffices to consider $\theta$, the unique non-zero component  of $%
\vec{\theta}$.

The transport of the winding angle is described by the average current density
\begin{equation}  \label{tok}
\vec{j}(\vec{x}) = \langle \theta \vec{v} \delta (\vec{x} - \vec{x}%
(t))\rangle,
\end{equation}
where the symbol $\langle \dots  \rangle$ denotes  averaging over a  set of Brownian trajectories, or, in the discrete
case,  of random walks.

In the case of an infinite plane (or a plane with periodic boundary conditions) and of
equilibrium, the particle density does not depend on the
coordinates and the average $\langle\theta\rangle$ vanishes identically  as
well as the average of the winding current density (\ref{tok}). If, however, a  confined planar geometry
is considered, then $\langle\theta\rangle$ will still
be vanishing but, as we will show below, this will not be the case for the winding current,
because of the geometrical constraints. Namely, due to the mere existence of  boundaries, there
will exist trajectories whose contribution to the winding will not be canceled
by that of other trajectories. 
As a result,
there will necessarily appear an anti-parallel edge current with $\theta > 0$ and $\theta < 0$
and $\vec{j}$ will not identically vanish anymore. These surviving edge currents can be
viewed as analogs of  spin edge currents in the spin Hall effect \cite{ZhouSpinHallCurrent}.

\section{Spin-orbit interaction and winding}

Before considering the winding currents, it is useful to discuss briefly an important
example of  spin-orbit interaction where the spin degree of freedom
depends on the winding angle $\theta$.

It seems known 
that in electronic transport in the presence of charged
impurities the electron spin depends on the winding of their trajectories.
This dependence manifests itself, for instance, as the so called \textit{skew
scattering} in the ``\textit{extrinsic}'' spin Hall effect \cite%
{RashbaExtrSpinHall,DasSarmaExtrSpinHall}. However, in the literature on the
subject, the concept of trajectory winding is mostly implicit.

Recall that in  electronic transport impurities interact with electrons and thus the
scattering angle, hence the deviation of the electron
trajectory from a ballistic straight line, depends, in relativistic theory, on the spin.
More generally, it is known that if, for any reason (not necessarily related
to electrons scattering on charged impurities) the trajectory of a relativistic
particle is not a straight line, the curving of the trajectory changes the
spin degree of freedom of the particle. This purely kinematic relativistic
effect is due to the successive non-parallel Lorentz transformations (%
\textit{boosts}) that lead to the so-called Wigner rotation of the spin, or the
Thomas-Wigner precession.

The variation of the precession angle of a classical rotating moment
resulting from an infinitesimal turning of the velocity has been obtained by
Thomas \cite{ThomasNature1926,jackson,To-Ok:97}
\begin{equation}  \label{eq:wigner_rotation}
d \vec{\varphi} \approx \frac{1}{2 c^2} \vec{v} \times d \vec{v} \ ,
\end{equation}
where $\vec{\varphi}$ is a vector of the precession angle, $\vec{v}$ is the
velocity, $d\vec{v}$ its variation, and $c$
is the speed of light.

Eq. (\ref{eq:wigner_rotation}) gives  the leading
relativistic approximation for the
precession angle.
It is worth mentioning that the exact relativistic expression is still a matter of controversy (see the recent articles \cite%
{Malykin2006, Ritus2011, ZanimovskyStepanovsky2010}). Note also that the
Dirac equation allows for a more precise study of this effect. However the
classical result (\ref{eq:wigner_rotation}) of Thomas is sufficiently accurate to describe  well  the
spin-orbit splitting in atoms \cite{Kessler}. It has played an important
role in the genesis of the quantum theory of radiation (see e.g \cite{To-Ok:97}).

Comparing (\ref{eq:thetadef}) and (\ref{eq:wigner_rotation}), we find a
simple relation between  the moment precession angle and the winding angle of
the trajectory
\begin{equation}  \label{eq:wigner_theta}
d \vec{\varphi} \approx \frac{v^2}{c^2} d\vec{\theta} \ ,
\end{equation}
where $d\vec{\theta}$ is the
trajectory turning angle. Note that, in many cases, the precession angle during the observation time
is essentially equal to the winding angle multiplied by a small parameter $v^2/c^2$
(e.g. the case of fermions at temperatures below Fermi temperature, or any
particles in the case of elastic scattering).

Let us now argue that the spatial curvature of a trajectory is equivalent to the
action of a magnetic field on a spin degree of freedom. Indeed, it is
known that an homogeneous magnetic field in the rest frame of a classical
spin results in its Larmor precession \cite{BargmannMichelTelegdi}
\begin{equation*}
\dot{\vec{s}} = \frac{g e}{2 m c} \vec{s} \times \vec{H} \ ,
\end{equation*}
where $\vec{s}$ is the spin in angular momentum units, $\vec{H}$ is the
magnetic field, $g$ is the gyro-magnetic factor, and $m$ is the
mass. Thus, an effective magnetic field with a Larmor frequency  equal
to the frequency $\dot{\vec{\varphi}}$ of Thomas-Wigner precession ($\dot{\vec{s}} = \vec{s}\times \dot{\vec{\varphi}} $) has the form
\begin{equation}  \label{eq-heff}
\vec{H}_{eff} = \frac{2 m c}{ge} \dot{\vec{\varphi}} \approx \frac{m}{e g c} \vec{v} \times \vec{\dot{v}} \equiv
\frac{2 m v^2}{e g c} \vec{\dot{\theta}} \ .
\end{equation}
The spin-field interaction energy is then
\begin{equation*}
E = \frac{g e}{2 m c}\vec{s}\cdot\vec{H_{eff}}=\frac{\vec{s}\cdot(\vec{v}%
\times \vec{\dot{v}})}{2 c^2} \equiv \frac{v^2}{c^2} \ \vec{s}\cdot \dot{\vec{\theta}} \ .
\end{equation*}
Hence, the direction and the magnitude of the effective
magnetic field are intimately related to the winding angle.

Returning to  random walks of electrons in solids (which can
result from electrons scattering on charged impurities, see e.g. \cite{Ak-Mo:07}), we
note that in this case it follows from (\ref{eq-heff}) that the effective
magnetic field of the Lorentz transformation of the impurity electric field
can be included in the Wigner-Thomas precession and vice versa, thus forming
a complete spin-orbit interaction.

We also note  that, leaving aside electrons in solids, the above discussion can
refer in general to the diffusion of any  particles with non-zero spin or magnetic
moment in disordered media, for example  spin 1 photons
whose motion in a medium with random refractive index can be described
as a diffusion process  \cite{Ak-Mo:07,Lo:12,Sh:06}), as well as in foams, metamaterials, dense
plasma, etc...(see e.g. \cite{Storzer2006} and references therein).

\section{Numerical simulations}

Let us now turn to  numerical simulations which do support the
existence of winding edge currents. We consider the
strip of width $L_{1}$ along the horizontal axis and of length $L_{2}$ along the vertical axis
on a planar square lattice of period 1.
In the longitudinal direction  periodic boundary conditions are imposed at $%
x_{2}=0,L_2$.
In the transverse  direction   reflecting (hard) walls enclose the strip in a finite width  $x_{1}=0,L_{1}$.

 We consider $N_{p}$ independent  particles whose trajectories are random walks on the lattice.
The diffusion time step $\tau $ is assumed  to be unity and  random hoppings  are allowed only to the nearest neighbor sites. Let
\begin{equation*}
\vec{x}(k,i)=\sum_{m=0}^{k}\Delta \vec{x}(m,i)
\end{equation*}%
and  $\theta(k,i)$ be respectively the position and the winding angle of the $i$th particle after $k$ jumps.  It is clear that each next turn (left or right) increases or
decreases $\theta (k,i)$ by $\pi /2$. We assume, for the sake of definiteness,
that a backward scattering does not change the winding angle of the
trajectory\footnote{%
We did consider  other variants, but they do not modify qualitatively the
effect, though it is rather obvious.}.

The time averaged winding current  $\vec{j} (\vec{x})$
at a point $\vec{x}$ of the lattice is after $N$ steps
(i.e. during the observation time $t = N$)
\begin{equation}
\vec{j}(\vec{x})= \frac{1}{N}\sum_{i}\sum_{k=0}^{N} \theta(k,i) \vec{v}(k,i)
\delta_{\vec{x}, \vec{x}(k,i)} \ ,  \label{eq:jpm}
\end{equation}
where $\delta_{\vec{x}, \vec{x^{\prime }}}$ is the Kronecker delta.

\begin{figure}[H]
\center
\subfigure[$\vec{j} (\vec{x})$]{\includegraphics[width=50mm]{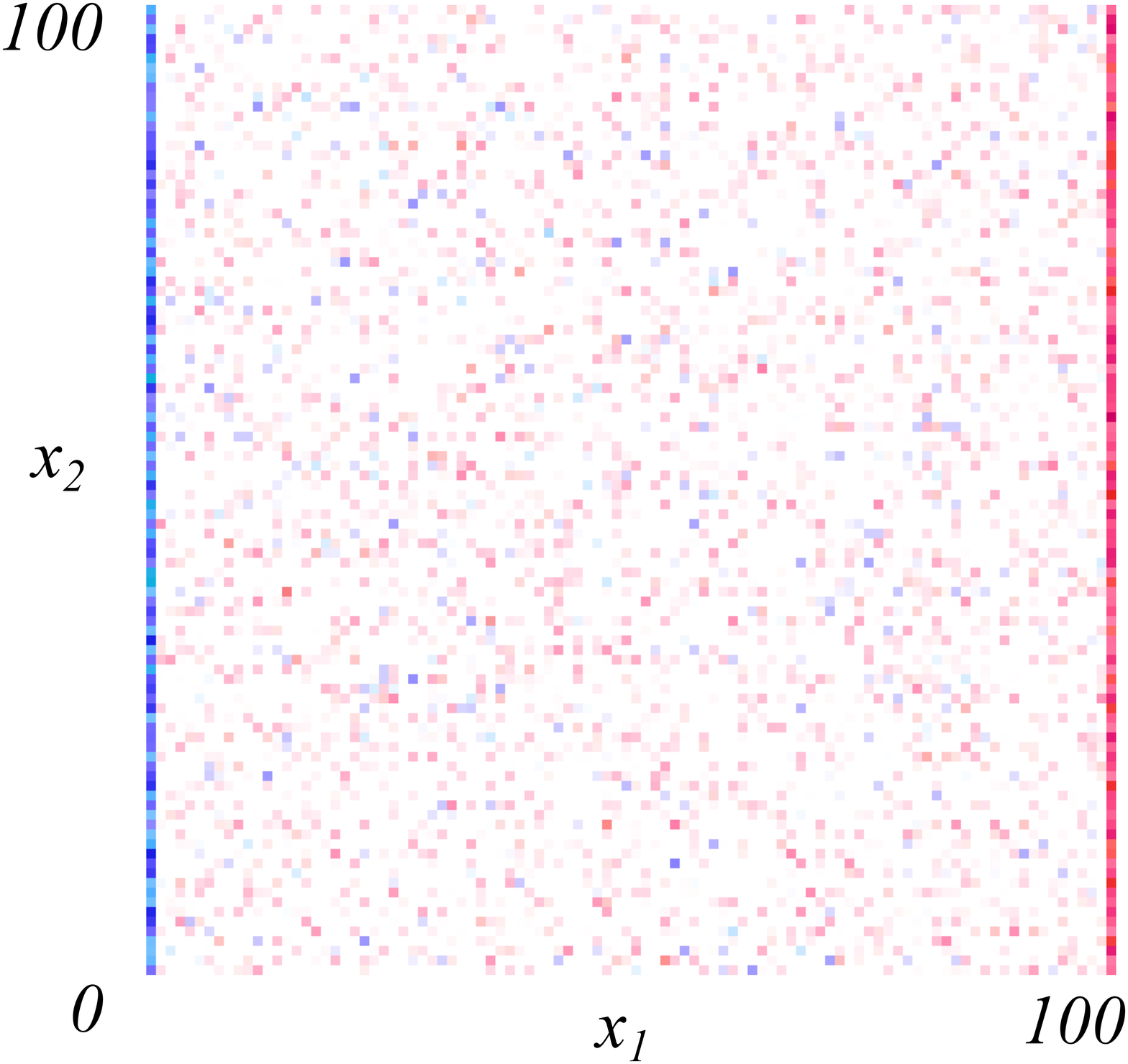}
\label{figure:jplus}}
{\subfigure[2d palette]{
\includegraphics[width=20mm]{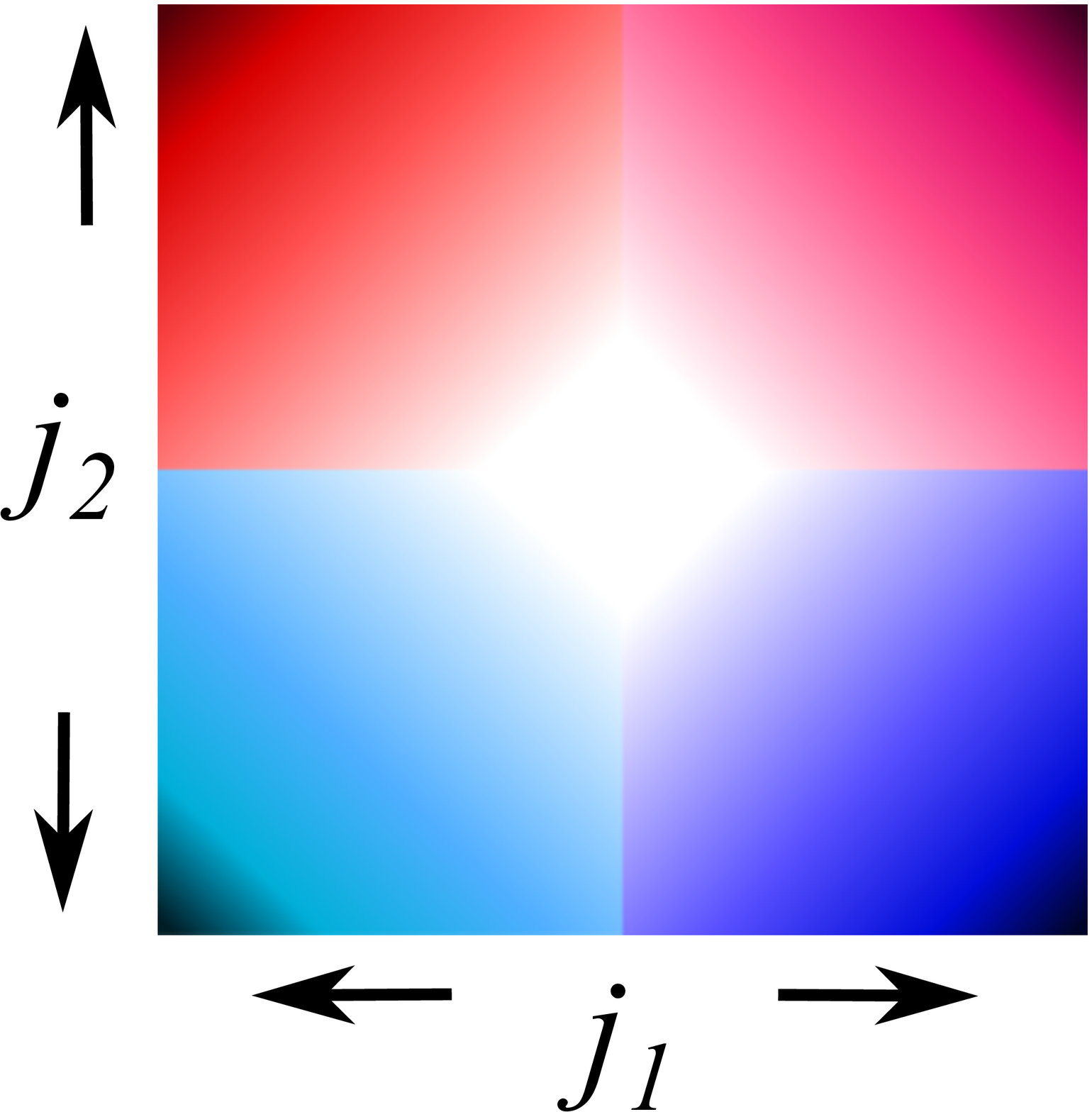}
\label{figure:palette}}}
{\subfigure[$L_2^{-1}\int^{L_2}_{0} dx_2 j_2(\vec{x_2})$]{\includegraphics[width=60mm]{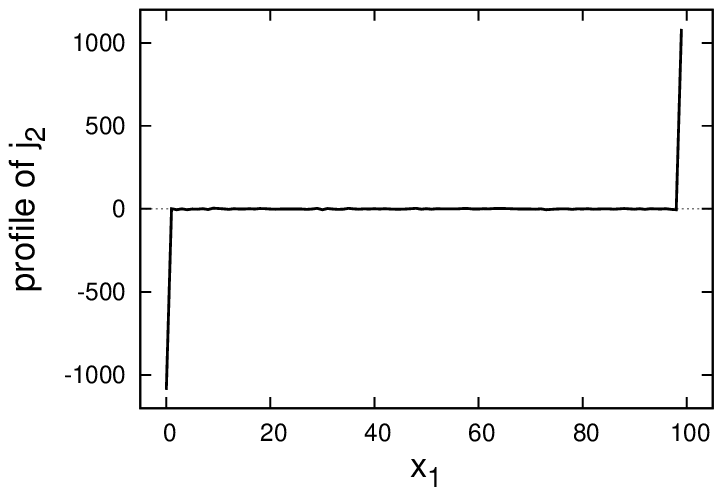}
\label{figure:jsum}}}
\caption{ \subref{figure:jplus} -- the winding current $\vec{j}$.
\subref{figure:palette} -- the 2d palette.
\subref{figure:jsum}
-- the winding current component $j_2$ averaged over $x_2 $ as a function of $%
x_1$.}
\label{figure:simulation-jvec}
\end{figure}

We analyzed a wide range of parameters $L_1$, $L_2$, $N$, and $N_p$. Typical
results are shown in Fig.\ref{figure:simulation-jvec} for $N_p = 4000$,
 $N=500000$, and  strip sizes $L_1=
L_2=100$. To display the 2d vector field $\vec{j}$, we use a 2d
palette as in Fig.\ref{figure:palette}. The palette center corresponds to zero
vector field and its vertical and horizontal edges correspond to the range
of the components $j_1$ and $j_2$ of $\vec{j}(\vec{x})$, respectively. 
The palette is divided in  four
sectors of different colors with a radial variation of the color
intensity. This allows to distinguish the approximate direction  and  magnitude of
the current.

The numerical simulations  for  $\vec{j}(\vec{x})$ are
shown in Fig.\ref{figure:jplus}.
In the bulk of the sample  only small-scale
fluctuations of the winding current can be seen. However, things differ strongly from the bulk in  
the narrow neighborhoods of the walls:  winding currents do exist  along the edges. They materialize  as almost continuous and
narrow lines on the left and right edges of the sample (blue and red,
according to the direction). The non vanishing longitudinal component of the winding
current, averaged over  $x_2 $, is displayed in
Fig.\ref{figure:jsum}. It
is concentrated within one lattice site from the
edges and directed downward near the left edge and upward near the right edge.

\begin{figure}[H]
\subfigure[$N = 5000$]{
\includegraphics[width=40mm]{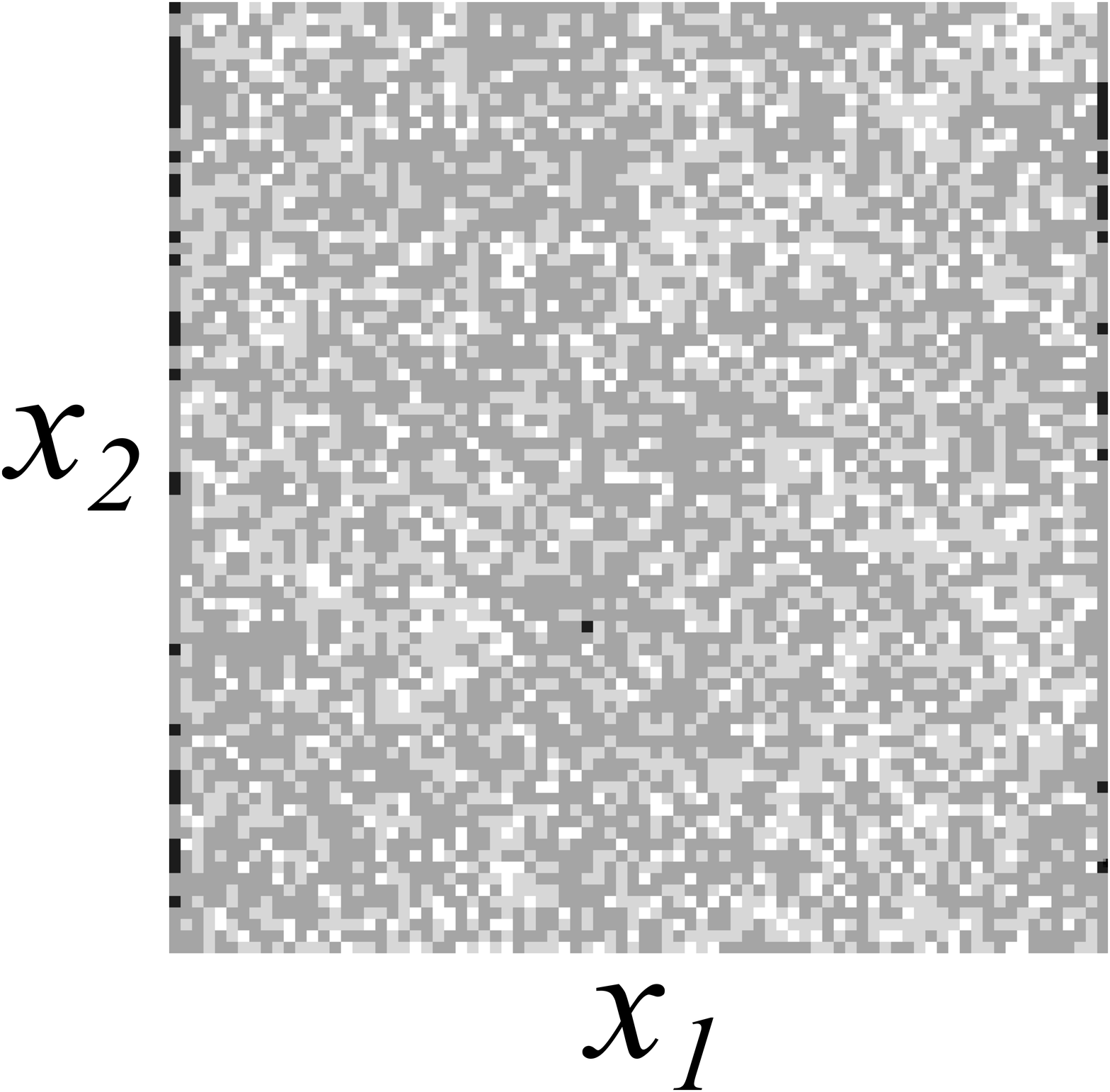}}
\subfigure[$N = 10000$]{
\includegraphics[width=40mm]{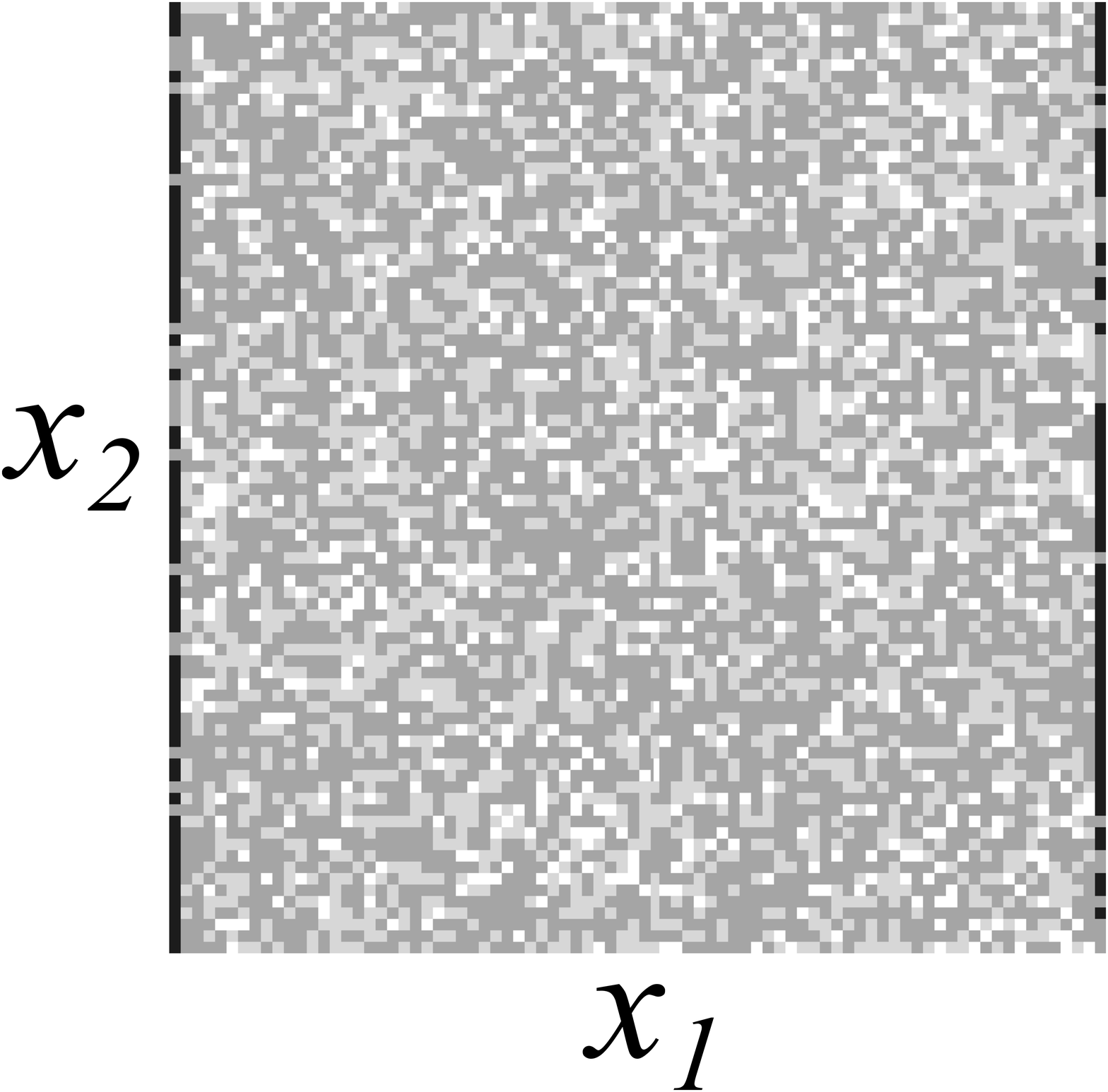}}
\subfigure[$N = 100000$]{
\includegraphics[width=40mm]{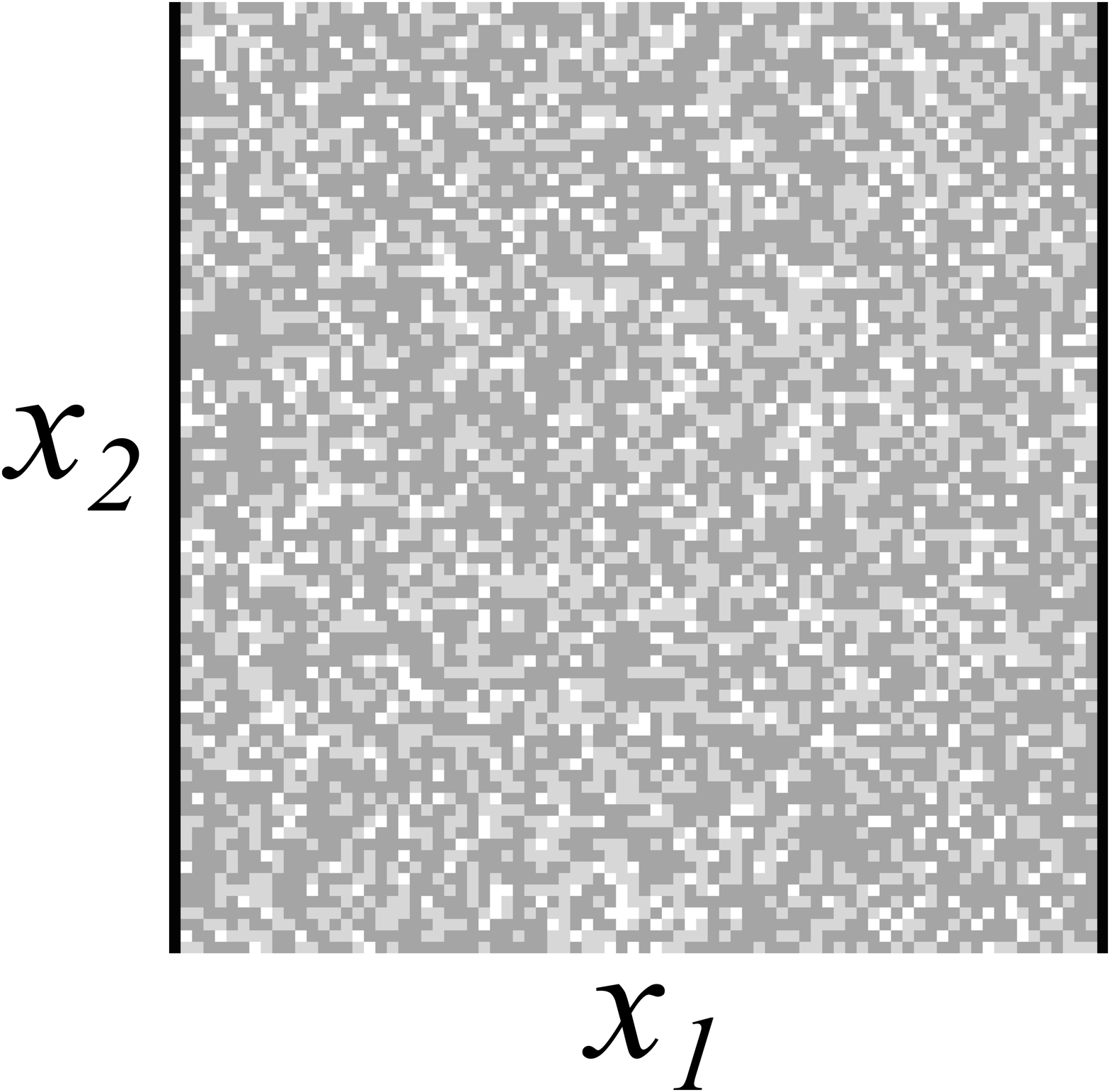}}
\caption{The modulus $|\vec{j}(\vec{x})|$ for different values of $N$ with $%
N_p = 3000$. The darker the dot, the bigger the modulus of  $\vec{%
j}$.}
\label{figure:jmodulus}
\end{figure}

Fig.\ref{figure:jmodulus} displays  the absolute value of the current density $|%
\vec{j}(\vec{x})|$ and how it concentrates on the edges with an increase of the simulation time $N$.

Since the system is in equilibrium, i.e., external fields are absent and the
particles are distributed uniformly,  winding currents can only originate from
the  confining boundaries and the chaotic kinetic energy of the
random walk  trajectories. Thus, these currents are by nature persistent.

\section{A simple solvable model}

Let us consider a simple model where the effect of winding currents is explicit. It consists of 
planar continuous Brownian trajectories described by an Ornstein-Uhlenbeck
process in a potential $U$ applied
along the $x_1$ axis
and infinitely growing as $|x_1| \to \infty$. The potential plays the role of the \textquotedblleft
\textit{soft walls}\textquotedblright\ confining  the Brownian particles,  and thereby encodes
the geometric constraints.
As in the previous section,  periodic boundary conditions at $x_{2}=0,L_{2}$ are assumed in the longitudinal direction.

The stochastic (Langevin) equations of motion are
\begin{eqnarray}\label{eq:langevin}
\dot{x_{i}}&=&v_{i} \\
\dot{v_{i}}&=&-\frac{1}{m}\frac{\partial U}{\partial x_{i}}-\beta
v_{i}+\xi _{i} \notag
\end{eqnarray}%
where $\vect{x}=(x_1,x_2)$ is the coordinate of the particle, $\vect{v}=(v_1,v_2)$ is its velocity, $m$ is the mass of the particle,
$\beta$ is the viscosity and $\vect{\xi}= (\xi_1,\xi_2)$ is the random force assumed to be the Gaussian
white noise, i.e.
\begin{equation}\label{wnc}
\langle \xi_i (t) \rangle=0, \quad \langle \xi_i (t') \xi_j (t'') \rangle = D \delta_{ij} \delta (t'-t'').
\end{equation}
Since we are interested in the winding of trajectories of the above random process, we consider the joint probability density of the coordinate, velocity and  winding angle $\theta$
\begin{equation}\label{distbis}
p(t,\vec{x},\vec{v},\theta )=\langle \delta (\vec{x}-\vec{x}(t))\delta (\vec{%
v}-\vec{v}(t))\delta (\theta -\theta (t))\rangle \ .
\end{equation}%
The $\theta $ integration of (\ref{distbis}) yields the density $P(t,%
\boldsymbol{x},\boldsymbol{v})$ satisfying the usual Kramers equation
\cite{VanKampen} (Fokker-Planck equation with external potential)%
\begin{equation}
\frac{\partial }{\partial t}P=-v_{i}\frac{\partial }{\partial x_{i}}P+
\frac{1}{m}\frac{\partial U}{\partial x_i} \frac{\partial }{\partial v_{i}}P+\beta \frac{%
\partial }{\partial v_{i}}v_{i}P+\frac{D}{2}\frac{\partial ^{2}}{\partial
v_{i}^{2}}P \ .  \label{kreq}
\end{equation}
where $D$ is defined in (\ref{wnc}) and the standard summation  convention  over
repeated indices is understood.

It is easy to check that the  probability  density $P$ in (\ref{kreq}) converges as $t \to \infty$  to a stationary
Maxwell-Boltzmann distribution
in $x_{1}$ and $\vec{v}$ with a  $L_{2}^{-1}$ normalization factor in  the longitudinal direction
$x_{2}$ (see (\ref{pst})). Note that one could also use reflecting boundary conditions or
a weak confining potential in this direction.

The Fokker-Planck equation for  $p(t,\vec{x}, \vec{v}, \theta)$
can be obtained as follows.
Let us denote
\begin{equation*}
\mathfrak{\Delta }=\delta (\vec{x}-\vec{x}(t))\delta (\vec{v}-\vec{v}(t))\delta (\theta -\theta (t)),
\end{equation*}
then $p=\langle \mathfrak{\Delta }\rangle $ and
\begin{equation*}
\frac{\partial }{\partial t}p=-\frac{\partial }{\partial x_{i}}\langle \dot{%
x_{i}}\mathfrak{\Delta }\rangle -\frac{\partial }{\partial v_{i}}\langle
\dot{v_{i}}\mathfrak{\Delta }\rangle -\frac{\partial }{\partial \theta }%
\langle \frac{\epsilon _{ij}v_{i}\dot{v_{j}}}{v^{2}}\mathfrak{\Delta }%
\rangle \ .
\end{equation*}%
Then  eq. (\ref{eq:langevin}) implies
\begin{multline}
\frac{\partial }{\partial t}p+v_{i}\frac{\partial }{\partial x_{i}}p =
\frac{\partial }{\partial v_{i}}\Big(\frac{1}{m}\frac{\partial U}{\partial x_{i}}
+\beta v_{i}\Big)p + \\ \frac{1}{mv^{2}}\epsilon _{ij}v_{i}\frac{\partial U}{\partial
x_{j}}\frac{\partial }{\partial \theta }p -\frac{1}{v^{2}}\Big[\frac{\partial }{\partial v_{i}}\delta _{ij}+\epsilon
_{ij}v_{i}\frac{\partial }{\partial \theta }\Big]\langle \xi _{j}\mathfrak{%
\Delta }\rangle \ .
\end{multline}%
To deal with the  correlation functions $\langle \xi _{j}\mathfrak{\Delta }\rangle $, we use the
Furutsu-Novikov formula (see e.g. \cite{LGP}, Chapter 10 and \cite{Kl:05},
Chapter 5) valid for any functional $R_{t}[\xi ]$ of the $\delta $-correlated Gaussian random
functions $\xi _{1}$ and $\xi _{2}$
\begin{equation*}
\langle \xi _{j}(t)R_{t}[\xi ]\rangle =\frac{D}{2}\Big\langle \frac{\delta }{%
\delta \xi _{j}(t)}R_{t}[\xi ]\Big\rangle \ .
\end{equation*}%
  We obtain after integration by parts
\begin{equation}
\langle \xi _{j}(t)\mathfrak{\Delta }\rangle =-\frac{D}{2}\frac{\partial }{%
\partial x_{k}}\Big\langle \frac{\delta x_{k}(t)}{\delta \xi _{j}(t)}\mathfrak{%
\Delta }\Big\rangle -\frac{D}{2}\frac{\partial }{\partial v_{k}}\Big\langle \frac{%
\delta v_{k}(t)}{\delta \xi _{j}(t)}\mathfrak{\Delta }\Big\rangle -\frac{D}{2}%
\frac{\partial }{\partial \theta }\Big\langle \frac{\delta \theta (t)}{\delta
\xi _{j}(t)}\mathfrak{\Delta }\Big\rangle \ .  \label{eq:maincor}
\end{equation}%
To find the variational derivatives of $x_{k}$ and $v_{k}$ with respect to $%
\xi_j $, we apply the operation $\delta /\delta \xi (\tau )$ to the integrated Langevin equations, then use the principle of causality and
take the limit $\tau \rightarrow t$. This yields
\begin{equation*}
\frac{\delta x_{k}(t)}{\delta \xi _{j}(\tau )}=\int\limits_{\tau }^{t}\frac{%
\delta v_{k}(t_{1})}{\delta \xi _{j}(\tau )}dt_{1}\xrightarrow{\tau
\rightarrow t}0\ ,
\end{equation*}%
and
\begin{equation*}
\frac{\delta v_{k}(t)}{\delta \xi _{j}(\tau )}=\delta _{kj}-\beta
\int\limits_{\tau }^{t}\frac{\delta v_{k}(t_{1})}{\delta \xi _{j}(\tau )}%
dt_{1}-\frac{1}{m}\int\limits_{\tau }^{t}\frac{\partial ^{2}U}{\partial
x_{k}\partial x_{i}}(\vec{x}(t_{1}))\frac{\delta x_{i}(t_{1})}{\delta \xi _{j}(\tau )}%
dt_{1}\xrightarrow{\tau \rightarrow t}\delta _{kj}\ .
\end{equation*}%
The variational derivative of $\theta $ in (\ref{eq:maincor}) is calculated
using (\ref{eq:thetadef})
\begin{equation*}
\frac{\delta \theta (t)}{\delta \xi _{j}(\tau )}=\frac{\delta }{\delta \xi
_{j}(\tau )}\epsilon _{ik}\int\limits_{0}^{t}dt_{1}\frac{v_{i}(t_{1})\Big(%
-\beta v_{k}(t_{1})-\displaystyle{\frac{1}{m}\frac{\partial U}{\partial x_{k}}(\vect{x}(t_{1}))}+\xi_k
(t_{1})\Big)}{v^{2}(t_{1})}\
\end{equation*}%
and finally
\begin{equation*}
\frac{\delta \theta (t)}{\delta \xi _{j}(\tau )}=\epsilon _{ik}\delta _{jk}%
\frac{v_{i}(\tau )}{v^{2}(\tau )}-\frac{\epsilon _{ik}}{m}\int\limits_{\tau
}^{t}dt_{1}\{...\}\xrightarrow{\tau \rightarrow t}-\epsilon _{ji}\frac{v_{i}(\tau )}{v^{2}(\tau )}\ .
\end{equation*}%
Thus, the correlation functions  $\langle \xi _{j}\mathfrak{\Delta }\rangle $ are expressed via the partial
derivatives of  $p$
\begin{equation*}
\langle \xi _{j}(t)\mathfrak{\Delta }\rangle =-\frac{D}{2}\frac{\partial }{%
\partial v_{j}}p+\frac{D}{2}\epsilon _{ji}\frac{v_{i}}{v^{2}}\frac{\partial
}{\partial \theta }p\ .
\end{equation*}%
We assume for the sake simplicity that the diffusion coefficient $D$ does
not depend on the velocity and we take into account the relation $\epsilon
_{ij}\epsilon _{jk}=-\delta _{ik}$. We obtain the Fokker-Plank equation
\begin{eqnarray*}
\frac{\partial }{\partial t}p &+&v_{i}\frac{\partial }{\partial x_{i}}p=%
\frac{\partial }{\partial v_{i}}\Big(\frac{1}{m}\frac{\partial U}{\partial
x_{i}}+\beta v_{i}\Big)p+\frac{D}{2}\frac{\partial ^{2}}{\partial v_{i}^{2}}%
p \\
&+&\frac{1}{m}\epsilon _{ij}\frac{v_{i}}{v^{2}}\frac{\partial U}{\partial
x_{j}}\frac{\partial }{\partial \theta }p+D\epsilon _{ij}\frac{v_{i}}{v^{2}}%
\frac{\partial ^{2}}{\partial v_{j}\partial \theta }p+\frac{D}{2v^{2}}\frac{%
\partial ^{2}}{\partial \theta ^{2}}p \ .
\end{eqnarray*}%
In the case at hand where the potential $U$ depends only on the transverse
coordinate $x_{1}$ (modeling the soft walls of a deep valley) the Fokker-Planck  equation becomes
\begin{eqnarray}\label{eq:kinetic}
\frac{\partial }{\partial t}p&+&v_{i}\frac{\partial }{\partial x_{i}}p-\frac{%
\partial }{\partial v_{1}}\frac{1}{m}U^{\prime }(x_{1})p-\beta \frac{%
\partial }{\partial v_{i}}v_{i}p-\frac{D}{2}\frac{\partial ^{2}}{\partial
v_{i}^{2}}p\\
&&=-\frac{1}{m}\frac{v_{2}}{v^{2}}U^{\prime }(x_{1})\frac{\partial }{\partial
\theta }p+D\frac{v_{1}}{v^{2}}\frac{\partial ^{2}}{\partial v_{2}\partial
\theta }p-D\frac{v_{2}}{v^{2}}\frac{\partial ^{2}}{\partial v_{1}\partial
\theta }p+\frac{D}{2}\frac{1}{v^{2}}\frac{\partial ^{2}}
{\partial \theta ^{2}}p \ . \notag
\end{eqnarray}%
Eq. (\ref{eq:kinetic})  differs from the Kramers equation by the terms on the
right-hand side. The physical meanings of the first and fourth terms
are the most interesting. It is already clear here and will
be even more clear below that the former is the current in the $x_{2}$ direction
and the latter is responsible for the diffusive spreading of $\theta $.

We also note that after integration over $\theta $ (\ref%
{eq:kinetic}) coincides with the Kramers equation (\ref{kreq}).

Eq. (\ref{eq:kinetic}) allows us to study the transfer of
arbitrary $\theta$-dependent quantities as well as non-equilibrium
situations corresponding to arbitrary initial conditions. In the particular case of the
winding current (\ref{tok}), which is linear in $\theta$, the steady-state
current can be found without actually solving the equation.


Indeed, it is clear that the transverse component $j_1$ of the mean current density  $\vec{%
j}$ vanishes. To find the average
longitudinal component $j_{2}$ let us multiply  (\ref{eq:kinetic}) by $%
\theta v_{2}$ and integrate over $\vec{v}$ and $\theta$.
We obtain after integration by parts
\begin{equation}
\frac{\partial j_{2}}{\partial t}=-\beta j_{2}+\frac{1}{m}U^{\prime
}(x_{1})\int d\vec{v}\frac{v_{2}^{2}}{v^{2}}P(t,\vec{x},\vec{v})\ .
\label{eq:jhydro}
\end{equation}%
The probability density $P(t,\vec{x},\vec{v})$ approaches for large time the stationary
distribution
\begin{equation}
P_{st}=\frac{\beta }{\pi DZL_{2}}\exp \left( -\frac{\beta }{D}v^{2}-\frac{%
2\beta }{mD}U(x_{1})\right) ,  \label{pst}
\end{equation}%
where
\begin{equation*}
Z=\int_{-\infty}^{+\infty} e^{- \frac{2 \beta}{mD}U(x_1)}dx_1
\end{equation*}%
is a normalization constant. It follows then from (\ref{eq:jhydro}) and (%
\ref{pst}) that for $t>>\beta ^{-1}$ the current $j_{2}$ relaxes to its stationary form
\begin{equation}
j_{2}(x_{1})=\frac{1}{2m\beta L_{2}}U^{\prime }(x_{1})P_{B}(x_{1})
\label{eq:mainresult}
\end{equation}%
where
$$
P_{B}(x_{1})\equiv \frac{1}{ Z}e^{-\frac{2\beta}{mD} U(x_{1})} 
$$
is the steady state (Boltzmann) probability distribution in the transverse direction.

We obtain finally for the local average density of  winding current per
unit length in the longitudinal direction%
\begin{equation}\label{cut}
J_{2}(x_{1})=\frac{n}{2m\beta }U^{\prime }(x_{1})P_{B}(x_{1}) \ ,
\end{equation}%
where $n=N_p/L_2$ is the number of particles per unit length in the longitudinal direction.

\begin{figure}[H]
\center
\subfigure[$U(x_1) = x_1^2$]{
\mbox{
\includegraphics[width=60mm]{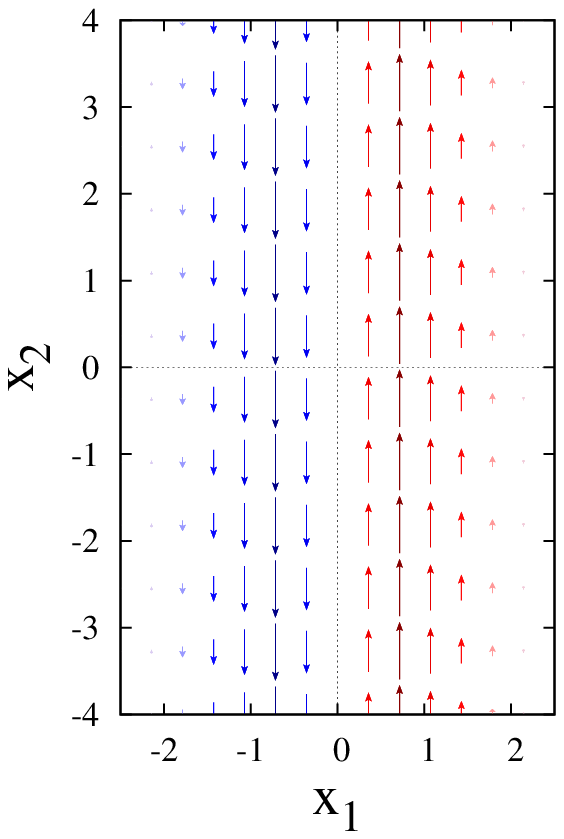} \hspace*{10pt}
\includegraphics[width=60mm]{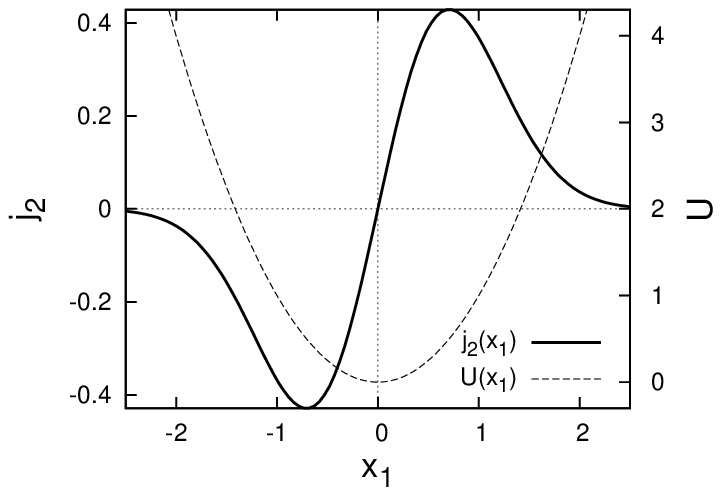} } }
\subfigure[$U(x_1) = \eta(-1-x_1)|x_1+1|+\eta(x_1-1) |x_1-1| $]{
\mbox{
\includegraphics[width=60mm]{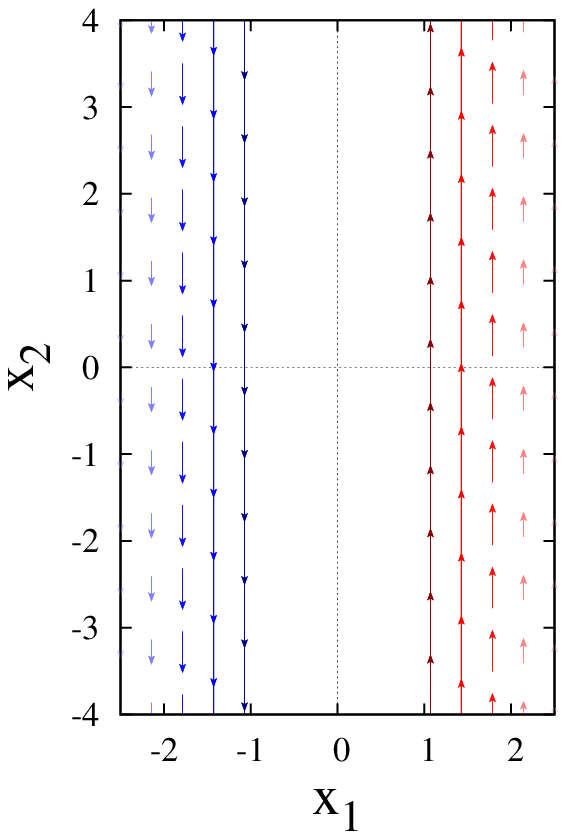} \hspace*{10pt}
\includegraphics[width=60mm]{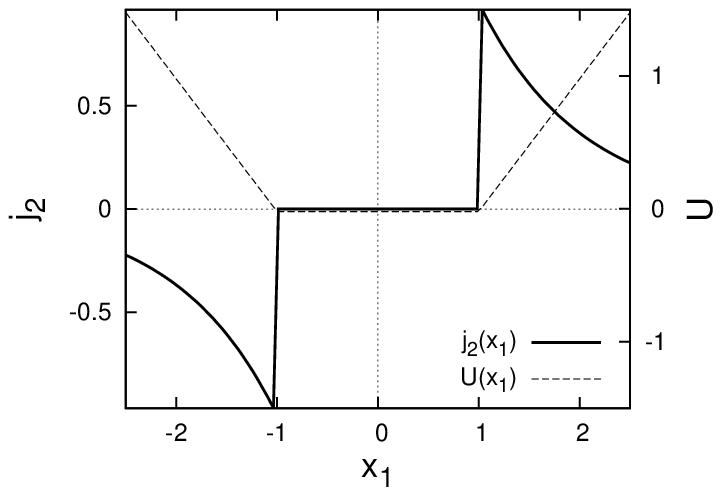} } }
\caption{The winding current density in the harmonic and trapezoidal
potentials and the profile $J_2(x_1)$ in arbitrary dimensionless units. On
the profile charts, the left y-axis is $J_2$ and the right y-axis is $U$.}
\label{figure:profile-linear}
\end{figure}

The average winding current density  (\ref{cut}) is proportional to the gradient of
the potential and to the local density of diffusing particles, which is
determined by the temperature and the height of the potential at a given
point $x_1$.  One  can easily see that the currents are antiparallel on  opposite walls.
The gradient of the potential and the Boltzmann distribution also determine
the currents width.
Particular examples of currents profiles are shown in Figure \ref{figure:profile-linear} for an
harmonic potential $U(x_{1})=x_{1}^{2}$ and  a trapezoidal potential
$U(x_{1})=\eta (-1-x_{1})|x_1+1|+\eta (x_{1}-1)|x_{1}-1|$, where $\eta(x)$ is
the Heaviside step function [ $\eta(x) = 1$ for $x > 0$, and $\eta(x) = 0$ for $x \leq 0$ ].

Note that formula (\ref{eq:mainresult}) can be interpreted
(up to a small relativistic factor) as  describing the diffusion of  magnetic
moments in an effective magnetic field $\vec{H}\propto \lbrack \vec{v}%
\times \nabla U]$, thereby corresponding to the setting of the spin Hall effect in the
potential $U$, as already mentioned above. We recall that in our approach there is no
interaction, and no magnetic (or any other type) moments. We are only dealing with 
distributions of trajectories winding, hence the above effect can be viewed as  purely dynamical
(\textquotedblleft winding Hall effect\textquotedblright ).

Assume finally without loss of generality that $U(0)=0$. Then the spatial average of $J_2$ in (\ref{cut})
over $(0,\infty)$ (or $(-\infty,0)$) is
\begin{equation}\label{mcur}
\int_{0}^\infty J_2(x_1)dx_1=\int_{-\infty}^0 J_2(x_1)dx_1=\frac{n D}{4 Z \beta^2}.
\end{equation}
This can be viewed as a manifestation of the weak dependence of our results on the
concrete form of the confining potential (considering that $n/Z$ is the effective planar density of particles).

\section{Conclusion}

We demonstrated that  random walks or  continuous Brownian motions in a confined planar geometry generate
dissipationless persistent currents of degrees of freedom associated with the
winding of their trajectories. This is a purely edge effect where the currents are
formed in a close neighborhood of the boundaries. In the case of the
\textquotedblleft soft walls\textquotedblright, where the geometric confinement  is
provided by an infinitely growing transverse potential, the channel width  of the
current is determined by the gradient of the potential  and the thermal energy of the
particles. In the case of reflecting boundaries (\textquotedblleft
hard-walls\textquotedblright ), the channel width is just
a single diffusion step, according to the numerical simulations.

In the microscopic world, the spin and geometric phase are  examples of
degrees of freedom which depend on the winding. Note that it is commonly
believed that in the spin Hall effect the dissipationless edge currents arise in the
conditions of the quantum \textquotedblleft \textit{intrinsic}\textquotedblright\
regime, where  dissipation and chaos are absent because of
quantization. Since, however,  the above \textquotedblleft
\textit{extrinsic}\textquotedblright\ spin Hall effect corresponds to an
ordinary diffusion, thus to a sufficiently high dissipation, one would
think that  persistent spin currents should not occur. However,
as we have shown, this is not the case simply because of the chaotic kinetic
energy of the random trajectories of the particles.

Similar effects may occur in purely mechanical systems where  relativistic or quantum
phenomena are absent. For example, the torque of a
compass arrow  in a moving car is also a degree of freedom which "feels" the curvature of the car trajectory. Thus, the  effect studied here can be of interest  beyond the scope of phenomena associated
with the spin Hall effect.

In the case of the Ornstein-Uhlenbeck model with soft walls we have obtained
an analytical expression for the persistent current of the winding via the
Boltzmann distribution in an arbitrary transverse confining potential. The kinetic
equation takes into account the winding of particles and, in principle,
allows to explore the joint probability distributions in the single-particle phase space
of the degrees of freedom with an arbitrary dependence on the winding.

\section{Acknowledgement}

This work was supported by the Ukrainian Branch of the French-Russian Poncelet Laboratory and the Grant 23/12-N of the National Academy of sciences of Ukraine.  
St\'ephane  Ouvry would like to thank the B. Verkin Institute for Low temperatures and Engineering for hospitality during the initial stages of the work.

\newpage

\textbf{LIST OF FIFGURES}

\bigskip

Figure 1.  
(a) -- the winding current $\vec{j}$. 
(b) -- the 2d palette. 
(c) -- the winding current component $j_2$ averaged over $x_2 $ as a function of $x_1$.

\bigskip

Figure 2.  The modulus $|\vec{j}(\vec{x})|$ for different values of $N$ with 
$N_p = 3000$. The darker the dot, the bigger the modulus of  $\vec{j}$.

\bigskip

Figure 3. The winding current density in the harmonic and trapezoidal
potentials and the profile $J_2(x_1)$ in arbitrary dimensionless units. On
the profile charts, the left y-axis is $J_2$ and the right y-axis is $U$.

\end{document}